\documentclass[letter]{aa} 
\usepackage{hyperref}
\hypersetup{colorlinks, allcolors=blue}
\usepackage{graphicx}    
\usepackage{color}  
\usepackage{txfonts}
\usepackage{inputenc}
\usepackage{multicol}
\usepackage{multirow}
\usepackage{longtable}
\usepackage{subcaption}  
\usepackage{siunitx}
\DeclareSIUnit{\year}{yr}
\DeclareSIUnit{\arcsec}{arcsec}
\usepackage{amsmath}
\usepackage{xcolor}
\usepackage{booktabs}
\usepackage{color} 

\usepackage{algorithm} 
\usepackage{algpseudocode} 
\begin{document} 

    \title{Direct Measurement of ISM Proper Motion with Image Registration}
    
   \author{M. Piecka\inst{1}
          \and
          L. Posch\inst{1}
          \and
          S. Meingast\inst{1}
          \and
          S. Hutschenreuter\inst{1}
          \and
          A. Rottensteiner\inst{1}
          \and
          J. Alves\inst{1}
          }

   \institute{University of Vienna, Department of Astrophysics,
              T\"urkenschanzstrasse 17, 1180 Vienna, Austria\\
              \email{martin.piecka@univie.ac.at}}

   \date{Received X XX, XXXX; accepted X XX, XXXX}


\abstract{

To date, quantification of the on-sky motion for interstellar clouds have relied on proxies such as young stellar objects (YSO) and masers. We present the first direct measurement of an interstellar cloud proper motion using the VISTA Star Formation Atlas (\mbox{VISIONS}) multi-epoch infrared images of the Corona Australis star-forming region. Proper motions are extracted by tracking the morphology of extended structures in the cloud complex based on image registration techniques implemented in SimpleITK. Our determined values ($\mu_{\alpha^*} \sim +15$ mas\,yr$^{-1}$, $\mu_{\delta} \sim -30$ mas\,yr$^{-1}$) are in good agreement with those obtained for YSOs and young stellar clusters in the region. This study demonstrates the potential of image registration for directly mapping the kinematics of nearby molecular clouds, opening a new window into the study of cloud dynamics. 

}

\keywords{ISM: clouds --
             ISM: kinematics and dynamics --
             ISM: structure --
             ISM: individual objects: Corona Australis}

\maketitle
%
\section{Introduction}\label{section:1}

Measurements of molecular cloud proper motions are critical for understanding how these stellar nurseries move, interact, and evolve. Molecular clouds are the main reservoirs of cold gas that fuel star formation, and their motion through the Galaxy offers key insights into Galactic structure and star formation processes. A cloud’s orbit and drift can reveal whether it follows Galactic rotation or displays peculiar motions driven by local disturbances. Investigating these motions helps to distinguish between different formation scenarios, such as triggered star formation, cloud–cloud collisions, or tidal shearing \citep[e.g.,][]{Elmegreen1998}.

Obtaining such measurements is very challenging. Typical molecular clouds in the solar neighbourhood span about ten parsecs and are seen (in the optical) as dark silhouettes against the background stellar field. Their patchy, diffuse, and extended structures lack the compactness required by traditional astrometric solution methods \citep[such as SExtractor combined with SCAMP,][]{SExtractor,SCAMP}. The proper motion of a molecular cloud is expected to be small (of the order of $\sim 10$ mas\,yr$^{-1}$ for nearby clouds), demanding high-precision data over long time baselines.

A direct measurement of the proper motion of a molecular cloud has long been considered virtually impossible. As a result, astronomers have relied on indirect methods, most commonly assuming that the average proper motion of young stellar objects (YSOs, especially embedded ones) reflects that of the parent cloud \citep[at least in a first order approximation, see][]{Ducourant2017}. Following the advent of {\it Gaia} astrometry \citep{GaiaDR1,GaiaDR2,GaiaEDR3,GaiaDR3}, large samples of young stars in star-forming regions now have precise proper motions, allowing researchers to derive bulk motions of the host clouds. For example, \citet{Grossschedl2021} investigated the 3D dynamics of the Orion cloud complex using {\it Gaia}~DR2 proper motions and ancillary radial velocities, revealing coherent radial gas motions at a 100-pc scale. This kinematic analysis suggests that the entire Orion cloud complex is responding to a significant feedback event that took place $\approx 6$~Myr ago \citep[see][and references therein]{Foley2023}.
Such analyses confirm that, to a good approximation, the mean space motion of very young stars can serve as a proxy for the motion of the molecular cloud that produced them.

Another valuable tracer of molecular cloud motion comes from maser astrometry in star-forming regions. Very Long Baseline Interferometry (VLBI) observations of water and methanol maser spots embedded in molecular clouds can yield extremely precise parallaxes and proper motions \citep{Sanna2010a,Sanna2010b}. Over the past decade, radio surveys have collected proper motion and parallax data for $\sim200$ masers associated with high-mass star-forming regions across the Galaxy \citep{Reid2019}, playing a key role in mapping Galactic structure.

\begin{figure*}[t!]
 \centering
 \includegraphics[width=2.0\columnwidth]{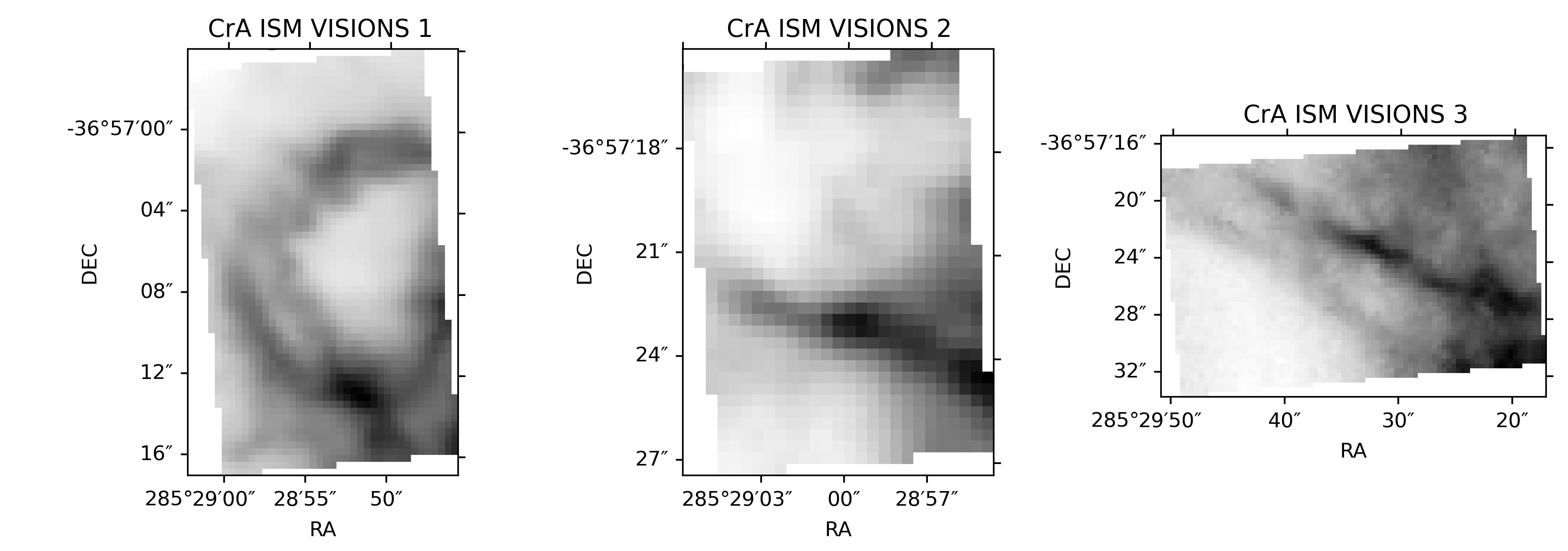}
 \caption{Negative images of the investigated CrA ISM regions. Based on the reference epoch $H$-band image from VISIONS (\texttt{CrA\_wide\_1\_4\_3\_A}). These images are reprojected and displayed in world coordinates -- the image registration was applied in pixel coordinates. No background removal procedure was used prior to registration.}
 \label{fig:masterA}
\end{figure*}

However, both proxy methods come with caveats. Not all young stars move in step with their parent clouds: internal velocity dispersion, stellar interactions, and the dynamical evolution of clusters can cause stars to drift relative to the gas. Massive stars can also perturb the surrounding interstellar medium (ISM) through radiation, winds, and supernovae, altering local gas kinematics \citep[e.g.,][]{Lopez_14_ApJ, Walch_15_MNRAS, Posch2025}. Similarly, masers trace only small, localised regions within molecular clouds (typically dense star-forming cores or shocked outflows) and their measured proper motions may include internal motions (e.g., outflows, infall) superimposed on the bulk cloud motion.

In this Letter, we present a direct measurement of the proper motion of the Corona Australis (CrA) molecular cloud, a nearby star-forming region at a distance of $\sim 150$~pc. Our approach relies on image registration methods widely used in medical imaging to determine the transformation mapping one image onto another. By comparing images taken several years apart, we detect angular offsets for nebulosities associated with the CrA cloud relative to the background sky. These offsets are interpreted as cloud proper motion.

\section{Image Data}\label{section:2}

\begin{table*}
\small
\caption{List of investigated objects and extracted proper motions, including clusters and YSOs with published astrometric data.}\label{table:1}
\centering
\begin{tabular}{l|rrrr|l}
\hline\hline
Object & \multicolumn{1}{c}{$\alpha$} & \multicolumn{1}{c}{$\delta$} & \multicolumn{1}{c}{$\mu_{\alpha^*}$} & \multicolumn{1}{c|}{$\mu_\delta$} & Source  \\
 & \multicolumn{1}{c}{[$^{\circ}$]} & \multicolumn{1}{c}{[$^{\circ}$]} & \multicolumn{1}{c}{[mas\,yr$^{-1}$]} & \multicolumn{1}{c|}{[mas\,yr$^{-1}$]} &  \\
  \hline
  
CrA~ISM~VISIONS~1 & 285.480917 & -36.951804 & 17.5 $\pm$ 0.7 & -34.5 $\pm$ 0.8 & VISIONS, this work  \\
CrA~ISM~VISIONS~2 & 285.483364 & -36.955935 & 16.8 $\pm$ 2.3 & -21.5 $\pm$ 1.3 & VISIONS, this work  \\
CrA~ISM~VISIONS~3 & 285.492770 & -36.957004 & 2.8 $\pm$ 5.3 & -30.6 $\pm$ 4.7 & VISIONS, this work  \\
\hline
HH~99B & 285.524000 & -36.910536 & 48.9 $\pm$ 4.4 & 1.9 $\pm$ 4.6 & VISIONS, this work  \\
CrA~ISM~VISIONS~HH & 285.552603 & -36.945413 & 73.2 $\pm$ 10.5 & -2.1 $\pm$ 17.2 & VISIONS, this work  \\
CHLT~15 & 285.494067 & -36.952587 & -2.6 $\pm$ 3.3 & -28.4 $\pm$ 5.1 & VISIONS, this work  \\
\hline
R~CrA & 285.4737 & -36.9524 & 7.6 $\pm$ 0.5 & -20.9 $\pm$ 0.5 & {\it Gaia}~DR3  \\
CrA-North & 280.1330 & -36.3063 & 0.7 $\pm$ 2.1 & -27.6 $\pm$ 1.2 & {\it Gaia}~DR3, \citet{Ratzenboeck_23a_AA}  \\
CrA-Main & 285.0759 & -36.9739 & 4.8 $\pm$ 1.0 & -27.2 $\pm$ 1.2 & {\it Gaia}~DR3, \citet{Ratzenboeck_23a_AA}  \\
CrA YSOs & 285.1298 & -37.1894 & 5.0 $\pm$ 1.0 & -27.4 $\pm$ 1.0 & {\it Gaia}~DR3, this work (Appendix~\ref{section:C})  \\
\hline
\end{tabular}
\tablefoot{Assuming a distance of 150~pc \citep[][and references therein]{Posch2023}, velocities in km\,s$^{-1}$ can be roughly estimated as $v_i = 0.711\,\mu_i$.}
\end{table*}

We take advantage of the capabilities provided by the VISIONS ESO Public Survey \citep[VISTA Star Formation Atlas; ESO program ID 198.C-2009;][]{VISIONS0, VISIONS1}, particularly its unique astrometric stability \citep{VISIONS2}. VISIONS is a near-infrared imaging survey that covers most of the major star-forming complexes within 500~pc of the Sun. It provides wide-field coverage (hundreds of square degrees) at sub-arcsecond resolution, with imaging depth sufficient to capture both faint stars and extended nebulosities. Crucially, VISIONS was designed as a multi-epoch survey: each region was observed several times over the 5~year survey duration. When combined with earlier data from the VISTA Hemisphere Survey \citep[VHS,][]{VHS}, it can achieve a time baseline of nearly a decade. Most importantly for our study, VISIONS provides absolute astrometry for all images with typical uncertainties below 10~mas in the astrometric calibration. While the survey’s primary goal is to trace the motions of young stars, it also captures the positions of associated nebular structures.

We make use of the following publicly available multi-epoch $H$-band images from the VISIONS survey to demonstrate the proper motion of the dust in CrA: \texttt{CrA\_wide\_1\_4\_3\_A} (2017-04-05) and \texttt{CrA\_deep\_wx} (2021-10-04). These are focused on the star-forming head of the cloud complex surrounding R~CrA (NGC~6729). To compare object positions, we used the header WCS information to reproject the second image onto the projection of the first using \texttt{reproject.reproject\_exact} \citep{reproject}. The pixel scale in the images was set to be 1/3\arcsec per pixel \citep{VISIONS2}.

Various ISM components can be readily recognised in CrA in near-IR extinction and starlight reflection, with the latter appearing more distinct in this region. While diffuse low-intensity structures can be seen in the images of CrA, our primary aim is to investigate well-defined, mostly narrow and bright ISM components, many of which can be identified within 1~arcminute from R~CrA. In this study, we focus on the following six objects (see Fig.~\ref{fig:masterA}, Fig.~\ref{fig:masterB}, Appendix~\ref{section:A}): an "S-shaped" filamentary structure (northern lobe CrA~ISM~VISIONS~1 and southern lobe CrA~ISM~VISIONS~2) located to the east of R~CrA, a lower intensity filamentary structure to the south-east from R~CrA (CrA~ISM~VISIONS~3), the YSO CHLT~15 \citep[or WMB~55, see][]{Choi2008}, the Herbig-Haro object HH~99B \citep[e.g.][]{2008A&A...481..123G}, and a previously unidentified Herbig-Haro object (CrA~ISM~VISIONS~HH). These objects were selected based on their reliable detectability in the images.

\section{Measuring Cloud Proper Motions}\label{section:3}

Compared to point source objects, the determination of the proper motion of extended ISM structures is a complex task. Assuming that an interstellar structure moves coherently across the sky, the primary objective of any method designed to tackle this challenge is the identification of structures and their morphological changes over time. In our case, this primarily entails the translational motion of ISM structures across different epochs. For this purpose, we utilised the SimpleITK Python package \citep{SimpleITK1,SimpleITK2,SimpleITK3}, which was developed for and is extensively used in medical imaging. SimpleITK is based on the Insight Toolkit \citep{ITK}, an open-source library of image processing tools. The image registration methods \citep[see][]{Maintz1998} are the primary tools of interest for our science case -- these procedures were designed to identify the transformations required to map an image onto a reference image. One of the main advantages of using image registration is the involvement of a metric that is used to assess the degree of similarity or alignment between images. This helps with structure identification between two images with differing brightness information, including systematic changes in the measured flux for extended objects due to variable background filters.

Our procedure for determining proper motions of ISM structures was designed as follows. Only sub-arcminute scale cutouts of the images (less than $100\,\textrm{px} \times 100\,\textrm{px}$) were selected for the analysis in order to reduce the influence of the internal motion. Based on our experiments with the SimpleITK setup (see Appendix~\ref{section:B}), we found that the accuracy of image registration is heavily affected by variable backgrounds (resulting mostly from differences in background modelling in the image processing pipeline) or unassociated ISM structures, stars, and bad pixels. To resolve this problem, background modelling and object masking are required to provide reliable measurements.
Unfortunately, the background structure in the CrA cloud is quite complex, and accurately accounting for the mentioned issues in the error estimation lies beyond the scope of this Letter.
The investigated Herbig-Haro objects and CHLT~15 represent exceptions to this problem, as they are compact and discernible from the surrounding ISM features, making it possible to model and subtract the background. We made use of the \texttt{scipy.ndimage.median\_filter} (with a filter size of 10~px) to represent the background model for both epochs.

In SimpleITK, we set up the registration method to make use of the \texttt{Correlation} metric for similarity measures (see Appendix~\ref{section:B}). Linear interpolation was used for estimating pixel intensity and the regular step gradient descent was selected as the optimiser. The transformation that we are interested in is pure translation, although more complex transformations can be included in the SimpleITK setup. 

To get the proper motion estimate of the targets, the pixel offset vector was determined with the registration method and converted to world coordinates using the WCS information included in the image headers. The projection-corrected proper motion in the right ascension, $\mu_{\alpha^*}$, was obtained by multiplying the calculated value by $\cos\delta$. Additional projection effects might affect the pixel information across the image, but these effects are negligible at these precision levels ($\gtrsim 1$~mas\,yr$^{-1}$) and spatial scales (less than $40\arcsec \times 30\arcsec$, $\delta \approx -37^{\circ}$).

SimpleITK image registration procedures do not offer any error estimates for the determined pixel offsets. To provide at least some uncertainty estimates, we assumed that the errors in flux can be described by a normal distribution. The regions surrounding the investigated Herbig-Haro objects were analysed to provide a background noise estimate. On the other hand, noise estimation within the CrA cloud is impossible without proper modelling of the ISM structure, forcing us to make use of the typical noise value across the image ($\approx 7$~ADU). We added the background noise and sampled the image fluxes over $n_{\textrm{samp}}=300$ iterations, determining the proper motion for each sample.

\begin{figure}
 \centering
 \includegraphics[width=\columnwidth]{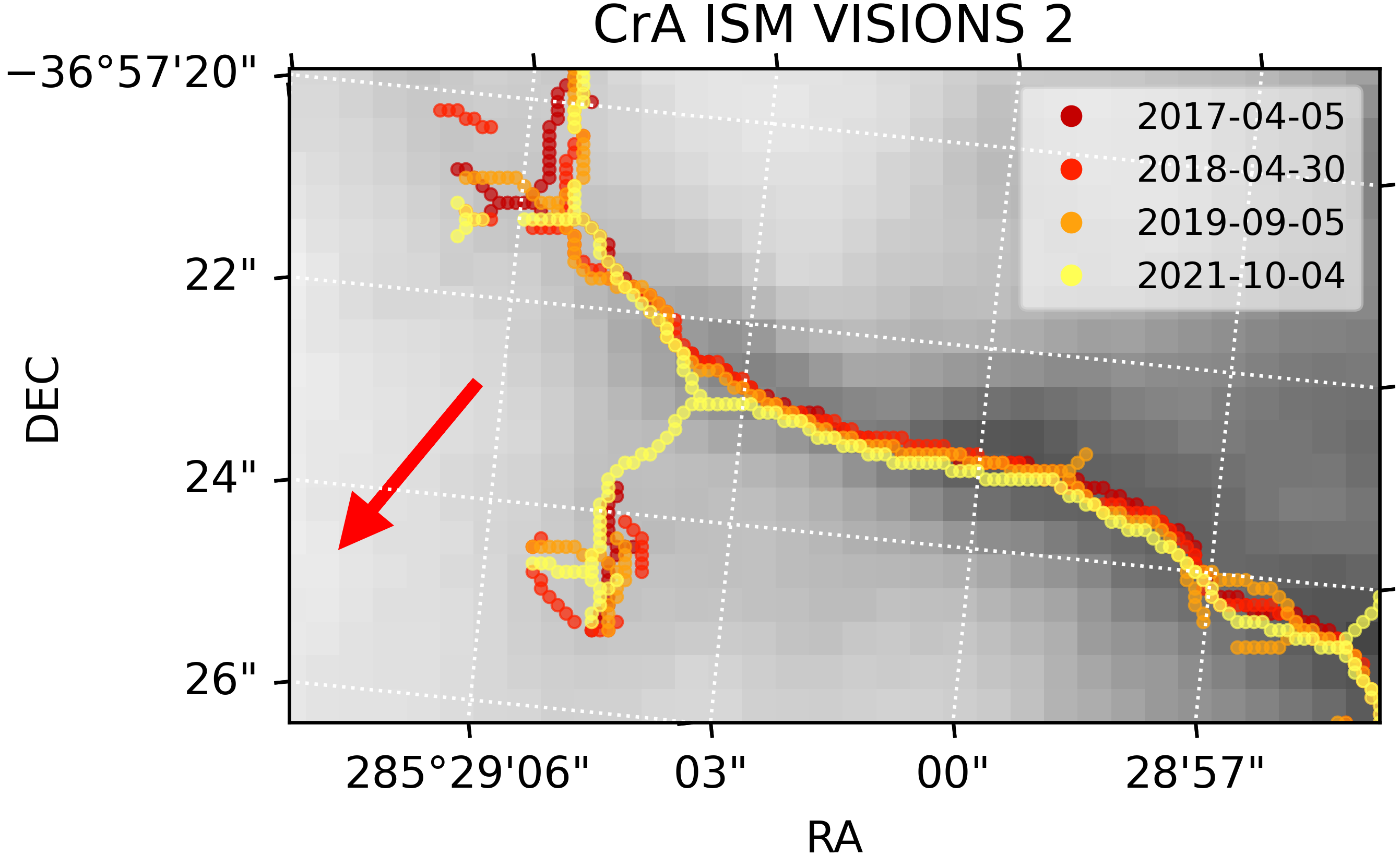}
 \caption{Visualisation of the gas proper motion for CrA ISM VISIONS 2. We computed the density peaks along the ISM structure (skeletons) across four epochs between 2017 and 2021 (including \texttt{CrA\_wide\_1\_4\_3\_C} and \texttt{CrA\_wide\_1\_4\_3\_F}), and overlaid them on the image from the first epoch. The red arrow shows the direction of motion derived using the methods described in Sect.~\ref{section:3}.
 }
 \label{fig:illustrate}
\end{figure}

The result of our procedure corresponded to distributions of proper motion components. The derived proper motions and their uncertainties for the investigated structures can be found in Table~\ref{table:1}, together with the mean motions of the clusters and the YSOs in CrA (see Appendix~\ref{section:C}).
A visualisation of the extracted proper motion can be seen in Fig.~\ref{fig:illustrate} -- the displayed image skeletons were obtained using \texttt{skimage.morphology.skeletonize}.
Animated images of the investigated regions (Sect.~\ref{section:2}) across four different epochs can be found at our online repository.\footnote{\href{https://github.com/mpiecka/ISM-Proper-Motions}{https://github.com/mpiecka/ISM-Proper-Motions}}

The influence of the parallax on the measured proper motions was ignored in this study. Its impact is inversely proportional to the time baseline. Based on the distance of the cloud and the time baseline, we estimate that neglecting parallax should have a minimal impact on our measurements, even in the worst case scenario ($\lesssim 1.5$~mas\,yr$^{-1}$).

\section{Discussion}\label{section:4}

Although the discrepancies in motion among individual ISM structures (Table~\ref{table:1}) are not fully explained by the estimated uncertainties, we note a very good agreement between the stellar motion and the mean of our measurements (see Appendix~\ref{section:B} for further details). This supports the validity of our assumed connection between the measured mean offsets in ISM positions and the CrA cloud's proper motion, opening up new options for studying cloud kinematics. Recent advances in astrometric surveys ({\it Gaia}, VISIONS) together with image registration methods show great promise in this endeavour, enabling first direct measurements of interstellar cloud proper motions.

While the determined values of the CrA proper motion seem comparable to those obtained for stars, we notice that the gas appears systematically faster, which might be explained by the cluster-chains scenario explored by \citet{Posch2023,Posch2025}. Alternatively, the provided uncertainties for the ISM regions CrA ISM VISIONS~1, CrA ISM VISIONS~2, and CrA ISM VISIONS~3 are most likely underestimated, contributing to the mentioned systematic difference. As discussed in Sect.~\ref{section:3} and Appendix~\ref{section:B}, the accuracy of image registration is heavily affected by the background structure of an astrophysical image. Since our error estimates are based purely on noise sampling, we ignore systematic effects, such as those resulting from the presence of the background. Future investigations of ISM proper motions will need to consider background structure modelling to improve the reliability of the results obtained with image registration methods. Additionally, while we showcase image registration as an option for determining the proper motion of a point source (CHLT~15), it is generally outperformed by procedures that track the position of sources' point-spread functions.

Since the observed cloud motion is based on the light reflected from ISM, and R~CrA shows a proper motion similar to that of the cloud, we should consider the case of a moving star illuminating a stationary cloud. An apparent motion of the cloud would be measured in such a configuration, with its magnitude depending on the relative distance between the ISM and the star. If the star is close to the ISM structure (the relative distance $d$ is comparable to the size of the structure $s$), the apparent motion of the ISM almost coincides with the motion of the star, although the illuminated shape of the ISM changes significantly over time. Alternatively, if the star is farther away ($d \gg s$), the illumination of the ISM is more isotropic, but the apparent motion of the ISM becomes smaller, scaling roughly as $\mu_{\textrm{app}} \propto d^{-1}$.

The images from VISIONS do not support the moving-star illumination scenario described in the previous paragraph -- systematic changes in the shape of the illuminated ISM are negligible. The shift in the position of the ISM structure appears systematic and independent of the position relative to the star.
Furthermore, all investigated ISM structures surrounding R~CrA display proper motions higher than that of the star. We conclude that the available observational data provide solid evidence of a true motion of the ISM. However, we do not rule out a possible minor systematic influence of the motion of the illuminating star on the measured motion of the ISM.

We highlight the potential of applying image-registration-based morphological tracking to other nearby clouds. Besides CrA, VISIONS has collected multi-epoch infrared images for other nearby molecular complexes (Ophiuchus, Lupus, Chamaeleon-Musca, Orion), many of which contain bright rim nebulae or filamentary structures that could serve as reference features. With the exception of Orion, the star clusters in these regions have significantly high proper motions \citep[$|\mu| > 20$~mas\,yr$^{-1}$,][]{Ratzenboeck_23a_AA} that should also be observed in the ISM, similar to the case of CrA. This makes the ISM associated with Sco-Cen the prime target for studying the full 3D kinematics of interstellar clouds.

Space missions capable of providing a higher astrometric precision than that of VISIONS would be ideal for making more precise measurements of proper motions and investigating structures at higher distances (such as Orion). The James Webb Space Telescope \citep[{\it JWST},][]{JWST} presently offers the best imaging capabilities for studying the proper motions of ISM structure with a resolution of 31 or 63 mas per pixel \citep[depending on the channel,][]{2022MNRAS.517..484N}, a factor of 5 to 10 times higher than in the case of VISIONS. Importantly, the images from {\it JWST} open the possibility to study motions of ISM not just in reflection but also in emission. We also note that an image of the $\rho$~Oph region was taken by {\it JWST}/NIRCam in March--April 2023. A second image taken in 2027 or 2028 should show a displacement of the cloud-associated ISM structures by $\sim 100$~mas or $\sim 2$~px, while only sub-pixel displacements can be investigated with VISIONS. Furthermore, Euclid and Nancy Grace Roman Space Telescope both aim at a high spatial resolution, approximately 2--3 times that of VISIONS \citep{Euclid_I,Roman}.

Ultimately, the ability to track morphologies of whole clouds over time will help constrain models of cloud formation and destruction, test feedback and turbulence theories, and refine our understanding of the link between gas kinematics and star-formation efficiency in our Galaxy. The presented work on CrA is a first step in this direction, demonstrating that the long-elusive proper motion of molecular clouds (and other ISM structures, such as Herbig-Haro objects) can be captured by tracking morphological features across different epochs.

\begin{acknowledgements}
Co-funded by the European Union (ERC, ISM-FLOW, 101055318). Views and opinions expressed are, however, those of the author(s) only and do not necessarily reflect those of the European Union or the European Research Council. Neither the European Union nor the granting authority can be held responsible for them.

Based on observations collected at the European Southern Observatory under ESO programme(s) 198.C-2009.

\end{acknowledgements}

\bibliographystyle{aa} 
\bibliography{aanda}

\begin{appendix}

\section{Selected objects in CrA}\label{section:A}

\begin{figure*}[t!]
 \centering
 \includegraphics[width=2.0\columnwidth]{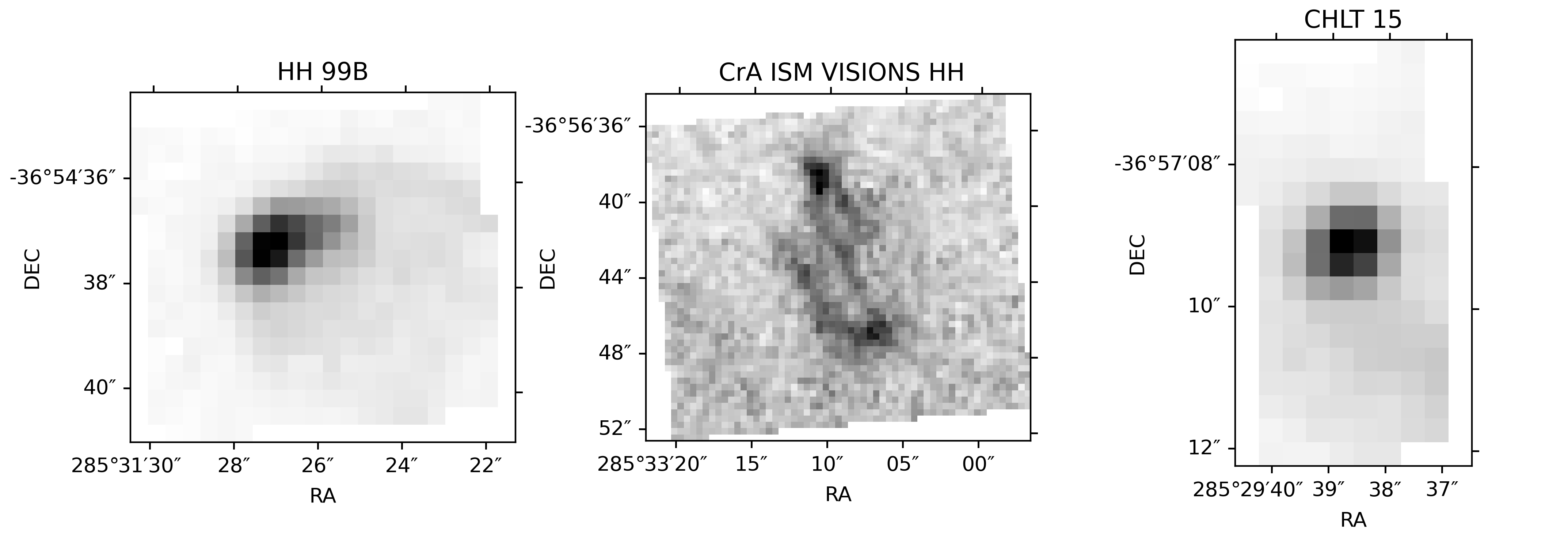}
 \caption{Negative images of the investigated Herbig-Haro objects and the YSO CHLT~15. Based on the reference epoch $H$-band image from VISIONS (\texttt{CrA\_wide\_1\_4\_3\_A}). These images are reprojected and displayed in world coordinates -- the image registration was applied in pixel coordinates. A simple background removal procedure was used prior to registration.}
 \label{fig:masterB}
\end{figure*}

The investigated "S-shaped" ISM structure is located $\approx 25\arcsec$ east of R~CrA. The centre of the shape is much brighter than its lobes and coincides with the position of FPM~13 (or CHLT~12), an extended emission source that was previously detected in radio observations \citep{Forbrich2006,Choi2008}. The lobes of the structure extend to the north (CrA~ISM~VISIONS~1) of this source and to its south (CrA~ISM~VISIONS~2). We note that the structure is most likely visible in the VISIONS images due to the dust reflection of starlight. The remaining ISM region, CrA~ISM~VISIONS~3, corresponds to low-intensity stripes located $\approx 60\arcsec$ south-east of R~CrA.

All three investigated ISM structures can be unambiguously identified in all three VISIONS bands. The YSO FPM~13 appears brighter at longer wavelengths (based on $J$-band, $H$-band, and $K_S$-band) when compared with the rest of the "S-shaped" structure. This is most likely caused by the relative absence of emission in the studied ISM structures. Furthermore, all ISM structures also appear in reflection when examining {\it HST} images \citep[F606W and F600LP,][]{HLA}.

Proper motion studies of Herbig-Haro objects have been conducted since the 1980s \citep[see][and references therein]{Reipurth2001}. To our knowledge, kinematics of such objects have not yet been investigated in CrA. Based on the results of image registration, we find that the directions of motion of HH~99B and the newly identified CrA~ISM~VISIONS~HH appear to be identical ($|\mu_{\alpha^*}| \gg |\mu_{\delta}|$), with velocity amplitudes that are relatively similar (within uncertainties). This might suggest that the two objects have the same origin. The large difference in their positions in declination would put their origin far to the west of R~CrA.

While the YSO CHLT~15 has already been mentioned in the literature \citep[e.g.][]{Choi2008}, its proper motion was never investigated. CHLT~15, also known as WMB~55 or SMM~2, is a Class~I object \citep[as determined by][]{Peterson2011}. Based on Table~\ref{table:1}, we find that the motion of this star matches the mean motion of the other YSOs in CrA. While image registration is not directly optimised for measuring proper motions of point sources, we can at least highlight the relatively high precision of our measurement for such a target, confirming that image registration serves as a very useful tool for measuring the motion of other (interstellar) objects in astronomical images.

Images of all CrA sub-regions investigated in this work are presented in Fig.~\ref{fig:masterA} and Fig.~\ref{fig:masterB}.

\section{SimpleITK performance test}\label{section:B}

\begin{figure}
 \centering
 \includegraphics[width=0.85\columnwidth]{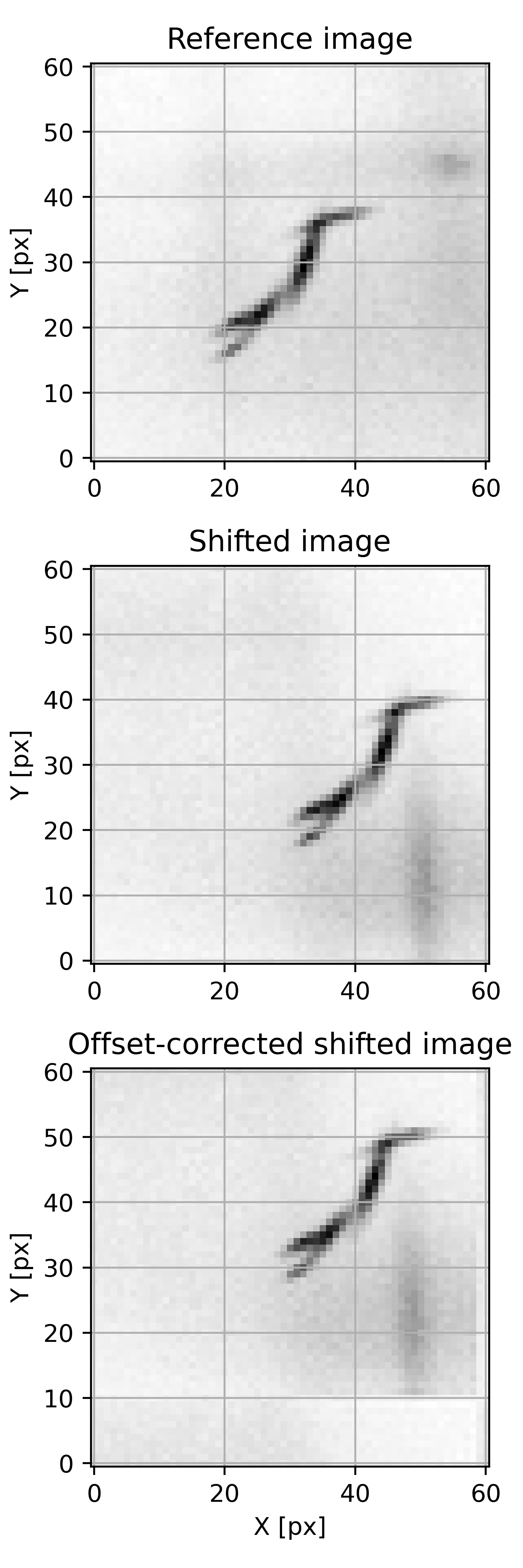}
 \caption{Extreme example of a failed registration for a case without background subtraction. This situation can occur even if the background model is identical in both images.}
 \label{fig:B1}
\end{figure}

To test the feasibility of SimpleITK for astrophysical cases, we prepared a simple procedure that generates synthetic structures in images (resembling ISM clouds). The structure itself consists of six 2D Gaussians with highly eccentric elliptical profiles, simulating the narrow shape of a filament. Secondly, a background image is generated that consists of ten randomised 2D Gaussians, representing complexity of background information and a possible occurrence of diffraction spikes and other artifacts. Finally, we add noise to the image ($\textrm{SNR} \sim 50$).

We create two images -- the reference image and a shifted image. The shifted image is generated by moving the simulated structure over a period of time at an extremely high velocity ($\mu_x = +100$ mas\,yr$^{-1}$, $\mu_y = +20$ mas\,yr$^{-1}$, $1\,\textrm{px} = 333\,\textrm{mas}$, $\Delta T = 40$ yr), resulting in significant offsets (a few pixels) that should be easily detectable. The noise and the background are computed and added separately for each "epoch", simulating complex changes in the image background. Afterwards, we make use of image registration as described in Section~\ref{section:3}, with a prior knowledge of the true motion.

\begin{figure}
 \centering
 \includegraphics[width=0.90\columnwidth]{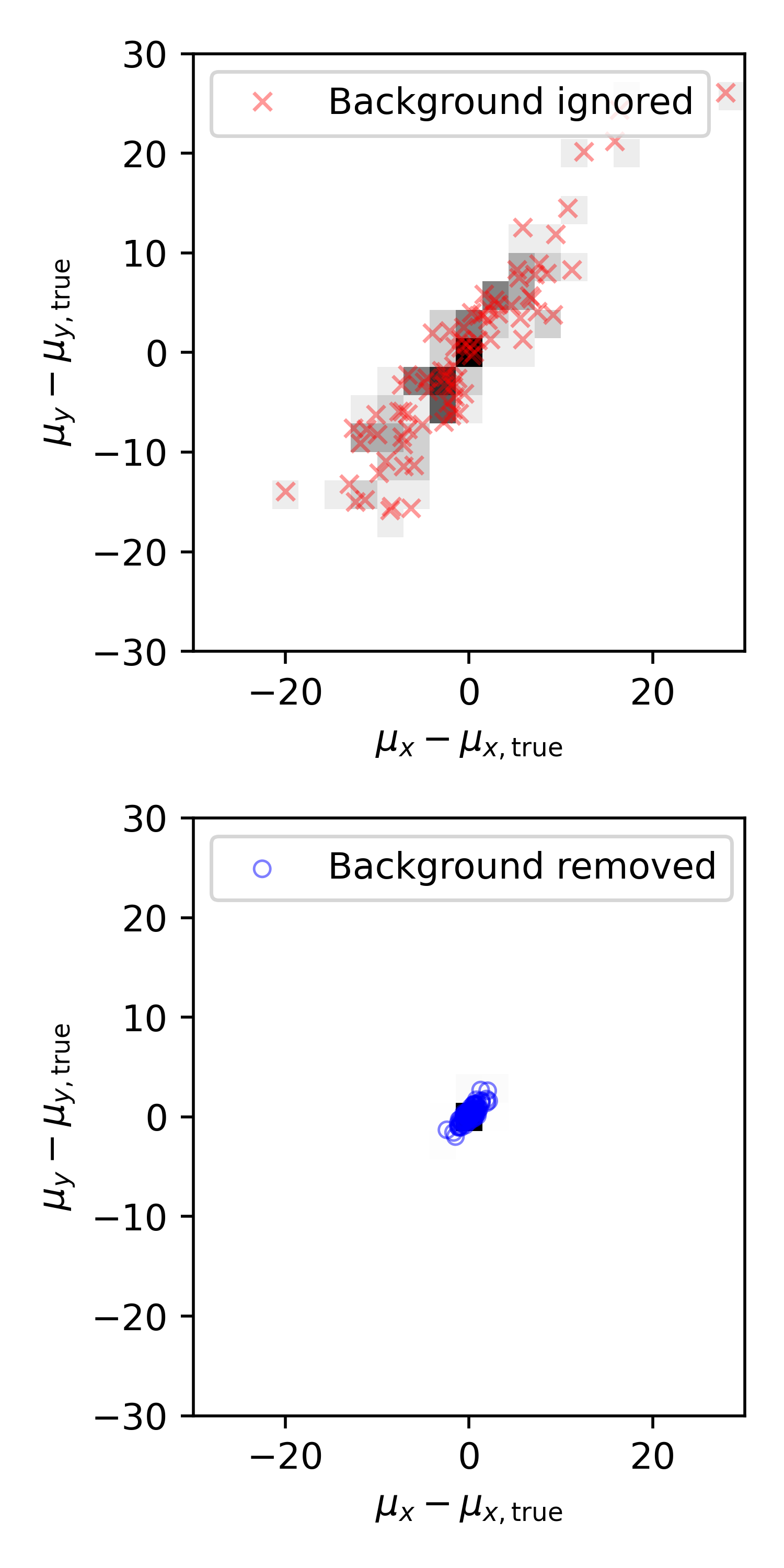}
 \caption{Comparison of extracted proper motions with true values from simulations of extreme offsets (see Appendix~\ref{section:B}). The simulation illustrates the difference between including and disregarding a background removal procedure with the presented method.}
 \label{fig:B2}
\end{figure}

The quality of the results depends heavily on the background information. The derived proper motion tends to be extremely imprecise if the background is not modelled (see Section~\ref{section:3}). One example of such failed registration can be seen in Fig.~\ref{fig:B1} -- the offset-corrected image does not match the reference image. Similar results are achieved if the same background structure is used in both images. We ran our synthetic generation code $n_{\textrm{sim}}=100$ times with and without our background modelling procedure ($n_{\textrm{samp}}=100$), and compared the results with the true proper motion values (Fig.~\ref{fig:B2}). We find that including even the simplest background modelling algorithm significantly improves the precision of the offsets extracted using image registration when a complex background is present in the image. Additionally, we see that the registration procedure generally results in a correlation between the extracted proper motion components. Although this effect depends on the morphology of the ISM structure and the direction of motion, it appears to be always present when using image registration (at least with our setup, see Section~\ref{section:3}).

We note that the presence of a bright spot (e.g., a star) in the background significantly affects the results of our measurements. Image registration might focus more on the motion of these bright spots (depending on the used similarity measure) and ignore the target of interest. If such a situation occurs, it would be important to mask the stars or to model and remove them from the images. In our investigation, we only encountered FPM~13 in the region of CrA ISM VISIONS~1. This source appears as an extended object and overlaps with the ISM features, making it difficult to remove. However, its intensity is not much higher than that of the surrounding ISM (peak flux ratio $\approx 1.5$). We estimated its motion to be $(\mu_{\alpha^*},\mu_{\delta}) \sim (0,-50) $~mas~yr$^{-1}$ by centring on the YSO and using a box of $24 \times 19$~px. Curiously, not including the section of the image with FPM~13 resulted in a proper motion of $(\mu_{\alpha^*},\mu_{\delta}) \sim (0,-45) $~mas~yr$^{-1}$ for CrA ISM VISIONS~1, suggesting that the influence of FPM~13 on the measurement in Table~\ref{table:1} is minimal.

The above mentioned experiments with synthetic and observed data indicate that our current image registration-based procedure lacks the ability to account for several key factors that influence proper motion measurements and their uncertainties. In particular, the choice of image region \citep[i.e. the box around the target, for a possible solution see][]{Reiter2017} can have a significant impact on the results, as demonstrated in the case of CrA ISM VISIONS~1. Using larger images in the SimpleITK may yield more robust results, but they are also more likely to contain various kinematically distinct ISM components, potentially introducing biases that are difficult to quantify.

It should be kept in mind that the corresponding offsets in VISIONS images are sub-pixel. A visual inspection alone cannot determine which of the derived proper motion values for CrA ISM VISIONS~1 is more reliable. Nevertheless, the direction of motion and the amplitude of the proper motion vectors in Table~\ref{table:1} are broadly consistent with the values obtained for the young stars. Further development of our procedure is necessary to enable reliable measurements of proper motions for individual ISM sub-structures.

Finally, the choice of the similarity metric in the SimpleITK setup is crucial for image registration. Both the \texttt{Demons} and the \texttt{MeanSquares} metrics rely on brightness consistency, a condition that is generally not satisfied when working with astronomical images. In contrast, the metrics based on mutual information or correlations are well suited for interstellar proper motions. Following the positive results reported by \citet{Lopez2022}, we decided to utilise the \texttt{Correlation}\footnote{See ITK Documentation for \href{https://docs.itk.org/projects/doxygen/en/stable/classitk_1_1CorrelationImageToImageMetricv4.html}{itk::CorrelationImageToImageMetricv4}.} metric available in SimpleITK. We note that the more sophisticated \texttt{ANTSNeighborhoodCorrelation}, which accounts for localised intensity patterns, should be considered in future investigations. In our case, the differences in the derived offsets between the two correlation-based metrics are negligible.

\section{Star clusters and YSOs in CrA}\label{section:C}

We compared the computed gas motion with that of the young stellar clusters associated with the CrA molecular cloud, based on the cluster census by \citet{Ratzenboeck_23a_AA}. Two clusters were identified in this region: CrA-Main (96 stars), projected directly onto the cloud, and CrA-North (351 stars), located closer towards the Galactic plane. Table~\ref{table:1} summarises the mean positions and proper motions of both clusters, with standard deviations given as errors. For these averages, we included all cluster members, as over 97\% (CrA-Main) and 98\% (CrA-North) have proper motion uncertainties below 0.5~mas\,yr$^{-1}$, with mean uncertainties of 0.1~mas\,yr$^{-1}$.

Following the method of \citet{Galli_20_AA}, who also identified these clusters, we selected Class~II YSOs within CrA-Main. Using the AllWISE \citep{WISE_10_AJ, NeoWISE_11_ApJ} based classification scheme from \citet{Koenig_Leisawitz_14_ApJ} ($W1-W2$ vs. $W2-W3$ colours), we identified 16 Class~II YSOs, likely the youngest stars formed from the CrA molecular cloud. Their average kinematic properties are also listed in Table~\ref{table:1}.

There appears to be an increase in proper motion from the CrA-North cluster to the Class II objects in CrA-Main, which continues with the motion of the cloud structure, particularly in CrA ISM VISIONS~1.

\end{appendix}

\end{document}